\documentclass[alpha-refs]{wiley-article}

\usepackage{graphicx}
\usepackage[space]{grffile}
\usepackage{latexsym}
\usepackage{textcomp}
\usepackage{longtable}
\usepackage{tabulary}
\usepackage{booktabs,array,multirow}
\usepackage{amsfonts,amsmath,amssymb}
\usepackage{natbib}
\usepackage{url}
\usepackage{hyperref}
\hypersetup{colorlinks=false,pdfborder={0 0 0}}
\usepackage{etoolbox}
\newif\iflatexml\latexmlfalse

\AtBeginDocument{\DeclareGraphicsExtensions{.pdf,.PDF,.eps,.EPS,.png,.PNG,.tif,.TIF,.jpg,.JPG,.jpeg,.JPEG}}

\usepackage[utf8]{inputenc}
\usepackage[english]{babel}

\usepackage{siunitx}
\usepackage{ulem}

\iflatexml

\else

\paperfield{Climate Dynamics}
\corraddress{Yunxiang Song, James Franck Institute, University of Chicago, Illinois, 60637, USA}
\corremail{songy1@uchicago.edu}
\fundinginfo{None.}

\papertype{Research Article}

\title{Excess Semiannual Variation in Historic Temperature Records}

\author[1]{Yunxiang Song}
\author[1]{Kyle B. Lawlor}
\author[1]{Thomas A. Witten}

\affil[1]{James Franck Institute, University of Chicago, Chicago, Illinois, USA}

\runningauthor{Yunxiang Song}
\begin{document}

\maketitle
\begin{abstract}
The annual temperature cycle of the earth closely follows the annual
cycle of solar flux. At temperate latitudes, both driving and response
cycles are well described by a strong annual sinusoidal component and a
non-vanishing semiannual component. A new analysis of historical weather station records in the United States determines persistent annual and semiannual variation with high precision. Historical annual temperature ranges are consistent with prior studies. Semiannual temperature cycles were much stronger than expected based on the semiannual solar
driving. Instead, these cycles were consistent with multiplicative effects of
two annual cycles. Our methods provide a quantitative window into
the climate's nonlinear response to solar driving, which is of potential
value in testing climate models.

\end{abstract}%

\section{Introduction}

Global climate features are often studied through terrestrial averaging of monthly mean temperatures over long periods of time. Several authoritative studies have revealed important global factors and secular trends \cite{legates1990mean, eliseev2003amplitude, stine2009changes, stine2012changes, mckinnon2013spatial}. Comprehensive grid maps of time-averaged and position-averaged temperature suggest good proportionality of the annual variation in temperature to that of the astronomical solar flux, or "insolation", especially in the northern mid-latitudes. Thus, describing the temperature response amounts to stating the relationship between the input and response sine waves. This inherently linear relationship can be characterized by gains and phase lags \cite{stine2009changes, stine2012changes, mckinnon2013spatial}. The gain is defined as the ratio of temperature variation to the corresponding insolation variation and describes the magnitude of response; the phase lag is defined as the difference between the phases of insolation and temperature cycles and gives the time delay of response.

The annual gain and the corresponding phase lags have been extensively used to show suggestive spatial and temporal features of the climate largely consistent throughout the northern hemisphere \cite{stine2009changes, laepple2009seasonal, stine2012changes, dwyer2012projected, mckinnon2013spatial}. Generally, terrestrial regions exhibit large temperature gains and short phase lags whereas maritime climates are characterized by small gains and long lags \cite{stine2009changes}. In addition to these well-documented findings, features of the yearly repetition show substructure that varies with geographic position. These variations have recently received much attention from climate scientists. The continuously changing Arctic is frequently linked with mid-latitude temperature amplifications \cite{overland2016nonlinear}. The time lag between maximum insolation and maximum temperature response in the spring was systematically observed to be different than that in the fall \cite{donohoe2020seasonal}. These results suggest the presence of nonlinear effects in the local driving-response system that are not well-understood. We provide an alternate approach to probe such nonlinearties by looking beyond the annual cycle.

Using harmonic analysis, we express the temperature or insolation over each day of a year as a sum of "annual" sine wave component plus a "semiannual" component with a half-year period, and so forth. We show a historically persistent substructure beyond the annual sinusoidal components: the harmonics of higher frequencies are well resolved above the noise. We quantify this persistence using a sample of records from the United States.

In this work, we focus on the semiannual temperature sine wave and the corresponding semiannual insolation, complementing prior studies of the annual sine-wave amplitudes. In the results section, we report anomalously large gains in the semiannual harmonic, which we show to drastically exceed their primary annual gains. This difference in gain across the semiannual and annual harmonics suggests a driving-response system incompatible with the simplest linear-response theory. In the discussion section, we interpret these semiannual temperature amplitudes in terms of a nonlinear process implicit in a standard description of the seasonal temperature variation.

\par\null

\section{Methods}
Our analysis below builds on the established understanding of factors affecting the seasonal temperature variation. It is known that the seasonal albedo, the local absorption of shortwave radiation, the re-radiated radiation, and the energy transported elsewhere along prevailing winds and temperature gradients can modulate the temperature and induce phase lags at any given location \cite{pierrehumbert2010principles}. Any historically persistent periodic variation seen on the earth is caused by the periodic cycle of insolation shaped by such time-invariant response properties of the earth. Gains and phase lags give a window into these time-invariant response properties. \cite{levitus1984annual, thomson1995seasons}.

We measured gains and phase lags for 64 weather stations spread over the United States, indicated by white dots as shown in Fig. 3. To improve precision, the stations were selected for consistency over long historic records of 50 to 130 years; to avoid urban effects, they were also selected for low population densities. To minimize short-term weather effects, we used the daily low temperature as our sampled quantity. In the interest of uniform data-taking, all stations were taken in a single country with institutional continuity. To ensure consistencies in gains and phase lags, we cross-checked our findings with an additional 9 weather stations scattered across the globe, selected on the basis of similar criteria. All annual temperature gains were similar to the well-established, uniform annual gains ($20$ \textdegree{}$C\cdot m^2\cdot W^{-1}$) across the mid-latitude oceans \cite{stine2009changes}. We express all annual gains henceforth as dimensionless quantities by normalizing to this “mid-ocean” gain.

\subsection{Analysis of data}
Our source of data was the National Oceanic and Atmospheric Association (NOAA) Global Historical Climatology Network dataset, using station temperature data from their first available year through 2010 \cite{menne2012overview, menne2012global}. We converted each calendar day in the records to fraction of an astronomical year, $t$, measured from the winter solstice. We averaged all the temperatures for a given $t$, for all years in each station record. The averaged temperatures for each station were then subject to a least-squares fitting to the function $T(t)$ given by
\begin{equation}
    T(t)=\bar{T}+a\cos(2\pi t)+b\sin(2\pi t)+c\cos(4\pi t)+d\sin(4\pi t)
\end{equation}
We determined the coefficients $\bar{T}$ and $a$ to $d$ using Mathematica’s LinearModelFit \cite{Mathematica}. We verified that the covariance matrix was very nearly diagonal, reflecting the orthogonality of the components in the fitting function. Given the values of these fit coefficients, we could readily determine the amplitude and phase $T_1$ and $\phi_1$ of the annual sinusoidal component and their counterparts $T_2$ and $\phi_2$ for the semiannual component. The temperature variation is thus given by
\begin{equation}
    T(t)=\bar{T}+T_1\cos(2\pi t-\phi_1)+T_2\cos(4\pi t-\phi_2).
\end{equation}
We compare this temperature variation at each station with the daily average insolation $S(\alpha,t)$ at that station's latitude $\alpha$ \cite{pierrehumbert2010principles}. We may express the known $S(\alpha,t)$ to good accuracy in the form
\begin{equation}
S(\alpha,t) = \bar{S}(\alpha)+S_1(\alpha)\cos(2\pi t)+S_2(\alpha)\cos(4\pi t).
\end{equation}

From the $T_n$ and $S_n$ we could compute the gains, $G_n$, for each harmonic. This gain is the temperature amplitude $T_n$ divided by the corresponding amplitude $S_1$ of the insolation $(G_n\equiv T_n/S_n)$. The insolation amplitudes all have a phase shift of $0$. Thus, the angles $\phi_n$ are the phase lags between the insolation $S_n\cos(2\pi nt)$ and the temperature $T_n\cos(2\pi nt - \phi_n)$. We report these gains and phase lags for two representative stations in Fig. 1. Table 1 shows the data for nine stations to indicate the variability of station data. Full results can be found in the supporting information. Their quoted uncertainties derive entirely from the standard errors in the fitted parameters. These uncertainties do not reflect the consistency of gains or phase lags over the historical record.

\subsection{Consistency of gains and phase lags}
To assess the variability of the annual phase lags over time, we performed a separate year-by-year analysis of the temperature records. For each station, we performed the five parameter fit (1) on the individual year data and obtained phase lags and annual and semiannual gains for each year. As a measure of consistency over time, we report the standard deviation in these individual year annual gains and phase lags as an additional uncertainty, presented in boldface in Table 1. The semiannual quantities for individual years were ill-defined and are not included.

\par\null

\section{Results}
The annual gains for 50 out of 64 stations were determined to within half a percent; the annual phase lags for all stations were determined to within half a day. Two such examples of the seasonal temperature cycle are given in Fig. 1(a). A complete depiction of the annual gain-phase distribution with location is given in Fig. 2. For most stations in the United States, the annual gains, normalized by the mid-ocean gain, varied across a range from 1 to 4, following a well-established geographic pattern \cite{stine2009changes}. The annual phase lags mostly converged within an angular range of 23.2 to 30.7 degrees (each degree is approximately a day), with no obvious geographic trends, as shown by the marker orientations in Fig. 2. The individual differences in annual gains and phase lags were typically of order ten times greater than the uncertainties arising from best-fit parameters of the seasonal cycles. 

To investigate the higher harmonics, we plot temperature profiles with the annual component and average removed in Fig. 1(b). The remaining semiannual signals are well resolved outside the noise. The error range of the pure semiannual fit is small, as represented by the orange band. Of all stations, 97\% had relative errors less than 25\% uncertainty in the semiannual gain.

The semiannual temperature amplitude relative to the corresponding annual amplitude is mapped in Fig. 3. The amplitude ratios are systematically smaller than 20 percent, with no clear latitude trends. One possible driver of the semiannual amplitude is the semiannual insolation amplitude $S_2$. To investigate the effect of $S_2$, we plotted the temperature amplitude ratio $T_2/T_1$ vs. the insolation amplitude ratio $S_2/S_1$ in Fig. 4. A $T_2$ driven entirely by $S_2$ would appear as a straight line through the origin on this plot. On the actual plot, no such straight-line correlation is apparent. Moreover, many gain ratios $G_2/G_1$, given by $(T_2/T_1)/(S_2/S_1)$, are as much as several hundred, implying large disparities between annual and semiannual gains (see supporting information).

\par\null

\section{Discussion}
Here we comment on our methodology and interpret the main findings reported above. Our nonstandard selection of data requires some justification. The daily low temperatures averaged over historical records at individual stations produce sinusoidal temperature cycles that agree well with the broad surveys of references \cite{stine2009changes, stine2012changes}. Further, our data reveal annual and semiannual variation with high precision. We acknowledge that our selection of daily low temperatures is not an optimal measure of mean temperature \cite{baker1975effect, schaal1977time}, though typically the temperature profiles constructed from daily low, daily high, or average daily temperatures barely depart from each other \cite{donohoe2020seasonal}.

Our data show a clustering of phase lags within a narrow band of $23.2$ to $30.7$ degrees. Our analysis of section 2.2 suggests that these lags are constant over time to within a few days, consistent with the observed station-to-station variability. The relative constancy of these lags over wide geographical regions was not noted in prior studies, though a similar narrow spread is visible in the supplementary Fig. S4b of Stine et al. (2009). We have added a band to Fig. 2 to indicate the range of their data for comparison.

Our annual temperature amplitudes of the annual sinusoidal component are in agreement in magnitude with prior studies. The annual component of temperature is certainly driven by the corresponding component of the insolation. Likewise, the semiannual temperature cycle is driven, to some degree, by the semiannual insolation. Yet, as Figs. 3 and 4 suggests, an explanation of the semiannual temperatures based on semiannual gain alone is incomplete. In the map of Fig. 3, the lack of latitude structure in $T_2/T_1$ indicates that the semiannual temperature amplitude $T_2$ is influenced by features additional to the latitude dependent semiannual insolation amplitude $S_2$. Similarly, the scatter plot of Fig. 4 suggests an apparent floor in semiannual relative temperature, of roughly $1.1$ times the relative insolation. This may indicate the contribution of semiannual gain as a baseline effect. The systematic departures from this baseline suggests that the bulk of the measured semiannual amplitude must come from another source. Nonlinear processes, such as two annual harmonics multiplying to produce a semiannual harmonic are a natural source. To show how two annual periodicities in climate features can lead to a semiannual temperature response, we focus on the relation between the insolation $S(\alpha, t)$ and the absorbed solar flux that is converted to heat, denoted $J(t)$. At any given station this $J(t)$ is smaller than $S(\alpha,t)$ because some of the incident radiation is re-radiated and not converted to heat. Thus $J(t)$ has the form $\beta(t)S(\alpha,t)$, where the location-dependent $\beta(t)$ is the absorbed fraction. the fraction $\beta(t)$ can depend on many climatic factors that vary in time and space. The seasonal variations of aerosols with complex refractive indices can directly scatter and absorb solar energy \cite{charlson1995aerosol}, and that of sub-micrometer aerosol particles can directly enhance both seasonal and spatial patterns of atmospheric albedo \cite{facchini1999cloud, mccoy2015natural}. Temporal variations in Planck, water vapor, and lapse rate feedback in the northern extratropics can impact the climate sensitivity to solar forcing \cite{armour2013time}. The efficacy of this forcing is modulated by the effective heat capacity of the atmosphere as well, which is determined by time-varying planetary boundary layer depth \cite{davy2016differences}. Over historic time, $\beta$ may be expressed in terms of location-dependent fourier components in the form $\beta(t)=\bar{\beta}(1+\hat{\beta_1}(S(\alpha,t)-\bar{S})+\mathcal{O}((S(\alpha,t)-\bar{S})^2))$. 

Here we demonstrate how this variation of $\beta$ can lead to semiannual periodicity in the temperature. To illustrate our point, we suppose that the induced variation is weak. Accordingly, we neglect any contribution beyond the linear term proportional to $\beta_1$. We keep only the annual sinusoidal modes of $\beta(t)$ so that $\beta(t)=\bar{\beta}(1+\beta_C\cos(2\pi t)+\beta_S\sin(2\pi t)+\text{higher frequency terms})$, where $\beta_C$ and $\beta_S$ are unknown coefficients. Further, for consistency with the form of $\beta(t)$, we factor out the annual average insolation $\bar{S}$ in equation (3) to obtain an equivalent representation with unit-free amplitudes $\hat{S_1} = S_1/\bar{S}$ and $\hat{S_2} = S_1/\bar{S}$ given by $S(\alpha,t) = \bar{S}\big(1+\hat{S_1}(\alpha)\cos(2\pi t)+\hat{S_2}(\alpha)\cos(4\pi t)\big)$. Hats hereafter indicate normalized quantities. Using these forms for $S(\alpha,t)$ and $\beta(t)$, we may readily find the time dependence of the absorbed flux $J(t)$.
\begin{equation}
    J(t)=\bar{\beta}\bar{S}\Big(1+\beta_C\cos(2\pi t)+\beta_S\sin(2\pi t)\Big)\Big(1+\hat{S_1}\cos(2\pi t)+\hat{S_2}\cos(4\pi t)\Big),
\end{equation}
Expanding this expression using trigonometric identities, we obtain
\begin{eqnarray}
  J(t) &= & \bar{\beta}\bar{S}\Big(\big[1+\tfrac{1}{2}(\beta_C \hat{S_1})\big]+\big[(\beta_C+\hat{S_1})\cos(2\pi t)+\beta_S\sin(2\pi t)\big] \nonumber \\
    && + \big[(\tfrac{1}{2}\beta_C \hat{S_1}+\hat{S_2})\cos(4\pi t)+\tfrac{1}{2}\beta_S \hat{S_1}\sin(4\pi t)\big] + \mathcal{O}(\beta_C \hat{S_2},\beta_S \hat{S_2})\Big),
\end{eqnarray}
where the first square bracket term is constant in time; the second has an annual period; and the third has a semiannual period.

The semiannual component has the expected part proportional to $\hat{S_2}$ from the semiannual part of the insolation $S$. In addition, it has a contribution proportional to $\hat{S_1}$ and to $\beta_C$ or $\beta_S$. Combining these annual and semiannual modes, $J(t)$ can be written 
\begin{equation}
    J(t)=\bar{J}\Big(1+\hat{J_1}\cos(2\pi t+\delta_1)+\hat{J_2}\cos(4\pi t+\delta_2)\Big),
\end{equation}
where $\hat{J_1},\delta_1,\hat{J_2},$ and $\delta_2$ can be expressed in terms of $\hat{S_1}, \hat{S_2}, \beta_S,$ and $\beta_C$.

We begin with the case without considering contributions from $\beta_C$ and $\beta_S$. Without the $\beta_C$ and $\beta_S$ factors, the temperature ratio $T_2/T_1$ is given by
\begin{equation}
    \frac{T_2}{T_1}=\frac{G_2}{G_1}\frac{S_2}{S_1}=\frac{G_2}{G_1}\frac{\hat{S_2}}{\hat{S_1}}.
\end{equation}
Here, the second equality makes the point that the ratio between un-hatted quantities is the same as that between hatted quantities. This formula predicts a simple proportionality between the $\hat{S_2}/\hat{S_1}$ and $T_2/T_1$, which are the quantities plotted in Fig. 4. Instead, the figure shows a broad scatter of points, implying $G_2/G_1$ values as high as several hundred. (The largest implied gains arise from latitudes near 44 degrees where the $\hat{S_2}$ goes to $0$.) The semiannual temperature variations are thus too large to be plausibly explained in terms of a linear gain treatment.

We now consider the effect of $\beta_C$, $\beta_S$ on the temperature $T(t)$ from (2). If the temperature response is proportional to the $J(t)$ with a frequency-dependent "heat gain" $\mathcal{G}_n\equiv T_n/J_n$ (in contrast with $G_n=T_n/S_n$) we then have $T_1=\mathcal{G}_1 J_1$ and $T_2=\mathcal{G}_2 J_2$, where $\mathcal{G}_1$ is the heat gain at the annual frequency and $\mathcal{G}_2$ is the heat gain at the semiannual frequency. Thus $T_2/T_1=(\mathcal{G}_2/\mathcal{G}_1)(J_2/J_1)$. The heat gain $\mathcal{G}$ gives the response of the steady-state earth system to an arbitrarily weak periodic driving. In general, such gains depend smoothly on frequency \cite{van2004detection}, so that $\mathcal{G}_2/\mathcal{G}_1$ would be a number of order unity. Any $\mathcal{G}_2/\mathcal{G}_1$ much larger or smaller than unity calls for an explanation. 

Including the effect of $\beta_C$ and $\beta_S$ on $J(t)$ in (5) gives a new contribution to the semiannual temperature $T_2$. From $(5)$, this new contribution dominates when $\beta_C\gg \hat{S_2} \hat{S_1}$. At latitudes where $\hat{S_2}$ vanishes, any nonzero $\beta_C$ part must dominate. The term in $\beta_S$ causes similar effects. To estimate the magnitude $\beta_C$ and $\beta_S$ needed to explain the semiannual temperature cycle, we neglect the $\hat{S_2}$ contribution to $J(t)$, so that the entire semiannual contribution comes from $\beta_C$ and $\beta_S$. For example, if only $\beta_C$ were present, (7) becomes 
\begin{equation}
    \frac{T_2}{T_1}=\frac{\mathcal{G}_2}{\mathcal{G}_1}\frac{\hat{J_2}}{\hat{J_1}}=\frac{\mathcal{G}_2\beta_C \hat{S_2}/2}{\mathcal{G}_1(\hat{S_1}+\beta_C)}=\frac{\mathcal{G}_2}{\mathcal{G}_1}\frac{\beta_C/2}{1+\beta_C/\hat{S_1}}.
\end{equation}
Here, we have used the fact that $J_2/J_1=\hat{J}_2/\hat{J}_1$. At latitudes near 44 degrees, the measured $T_2/T_1$ ratios from the full data table lie in the range of 5-10 percent (supporting information). We readily verify that this range is compatible with $(8)$ for $\mathcal{G}_2/\mathcal{G}_1\approx 1$ and $0.1\lesssim\beta_C\lesssim 0.3$. Other latitudes are also consistent with this range. This amount of periodic annual variation is consistent with other evidence, such as the atmospheric coalbedo. 

We now consider possible physical sources of the $\beta_C$ and $\beta_S$ variation from among the many possibilities noted above. One documented contributor is the annual variation of the coalbedo \cite{kukla1980annual}. For more recent years, the spatially-resolved variation of top-of-atmosphere coalbedo corroborates these findings. This contribution appears sufficient to account for the anomalous $T_2/T_1$ as discussed in the Appendix.

This example shows that the primary insolation cycle when combined with the induced annual variation in some other quantity, such as the coalbedo, can plausibly account for our observed semiannual amplitudes in temperature. Other time-dependent cycles of similar magnitude would have the same effect. Our observations are thus compatible with the many potential causes stated above. Without further study, the specific nonlinear effect responsible for the part of $T_2$ beyond baseline solar driving is not clear. However, the existence of this semiannual variation sets a lower bound on the importance of these nonlinear effects. Thus it provides an alternate route to proving the sensitivity of properties like the albedo to the insolation. Further, our study gives strong evidence against the simple picture wherein the semiannual amplitude $T_2$ is simply caused by the corresponding solar amplitude $S_2$.

To state the temperature gain for a particular harmonic as a simple ratio between temperature and insolation amplitudes is therefore incomplete. In terms of the atmospheric coalbedo, a more direct way to quantify the response of temperature to solar forcing in the $n^{th}$ harmonic is to take $\mathcal{G}_n=T_n/J_n$, where $J_n$ is the energy flux after interference is taken into account. More broadly, for further research that studies surface air temperatures in response to external solar forcing, this modified definition of gain can be a more accurate measure to understand how the effect of this forcing varies over frequency components of the temperature. We note that the annual amplitudes of $\beta(t)$ ($\beta_C, \beta_S$) appear in both the annual and semiannual components of Eq. 5, thus providing a redundant measure of $\beta_C$ and $\beta_S$.  The annual part of Eq. 5 appears compatible with observations about the asymmetric maxima and minima of the annual temperature cycle \cite{donohoe2020seasonal}.

\section{Conclusion}
The analysis of gains and phases of the annual temperature cycle has proven to play a substantial role in understanding the underlying mechanisms that govern the earth's thermal response to insolation \cite{stine2009changes, stine2012changes}. A growing number of recent studies have begun to probe residual features of the annual cycle beyond the fundamental annual sine wave \cite{armour2013time, overland2016nonlinear, donohoe2020seasonal}. We showed that gain-phase analysis incorporating higher harmonics of the temperature variation revealed anomalously large semiannual gains that are ill-understood under the conventional assumption of linear-response to $S(\alpha, t)$. We pointed out a feature of a widely accepted theoretical account of seasonal temperature in terms of interference of annual components as a candidate for explaining this anomaly. For example, the insolation forcing can interact with an annually varying coalbedo to produce extra signal leaking into the higher harmonics.

Extensions of this study beyond its limited geographical scope show promise in distinguishing baseline behavior from nonlinear forcing, and in identifying extents of nonlinear driving and response in the climate cycle both regionally and globally.

\section{Appendix}
Here, we take the absorption factor denoted by $\beta(t)$ in the main text to be one of the many plausible sources, namely the atmospheric coalbedo. We provide recent evidence from albedo data to support our estimates for the one-year sinusoidal amplitude of $\beta(t)$. This data is derived from the CERES-SSF1deg dataset reporting top-of-atmosphere albedo data from satellite observations \cite{doelling2013geostationary, doelling2016advances}. We restricted grid bounds to encompass the United States, spanning longitudes 55 - 120W and latitudes 25 - 50N. We subtracted the gridded-values of albedo from one to obtain the coalbedo, and selected for years from the first available year (2000) to the last year in our NOAA temperature records (2010) \cite{menne2012global, menne2012overview}. This complements the evidence presented in Kukla \& Robinson (1980) by providing more recent observations in the coalbedo time variation to cross-check with our magnitude feasibility argument presented in the discussion section. Monthly coalbedo values are plotted for one particular location in Fig. A1. We first adjusted the coalbedo variation relative to 1 instead of its annual average by factoring out this average. We then fitted these values to a sinusoidal function with period a year and obtained a dimensionless amplitude $\beta_C$ that is compatible in magnitude with our discussion. We plotted these annual amplitudes for all grid locations across the United States in Fig. A2. These amplitudes lie mostly in the range of 0.1 - 0.4, consistent with our estimates of $\beta_C$ in the main text. We have not attempted a more refined analysis in this initial study.

\section{Acknowledgements}

The authors thank professors Noboru Nakamura, Mary Silber, Sidney Nagel, Dorian Abbot, and Robert Rosner for constructive discussions and valuable comments on the manuscript; professors Alexander Stine and Peter Huybers provided useful insights during the early stages of the work. We thank Aaron Donahoe for careful reading of an earlier version, leading to multiple improvements. The primary temperature data used in this paper can be found by accessing the National Oceanic and Atmospheric Association (NOAA) Global Historical Climatology Network dataset. The top-of-atmosphere albedo data used in the appendix can be found by accessing the NASA CERES Single Scanner Footprint (SSF1deg) dataset. Yunxiang Song acknowledges support from a James Franck Institute summer fellowship.

\section{Conflict of interest}

The authors declare no conflict of interest.

\section{Supporting Information}
The supporting information contains a derivation of the annual and semiannual insolation amplitudes as coefficients of a Fourier series and the full data table. It also contains graphs of full temperature profiles and semiannual temperature cycles for all stations.

\selectlanguage{english}
\bibliography{Song_refs.bib}

\raggedbottom

\pagebreak

\begin{figure}
    \centering
    \includegraphics[width=0.70\columnwidth]{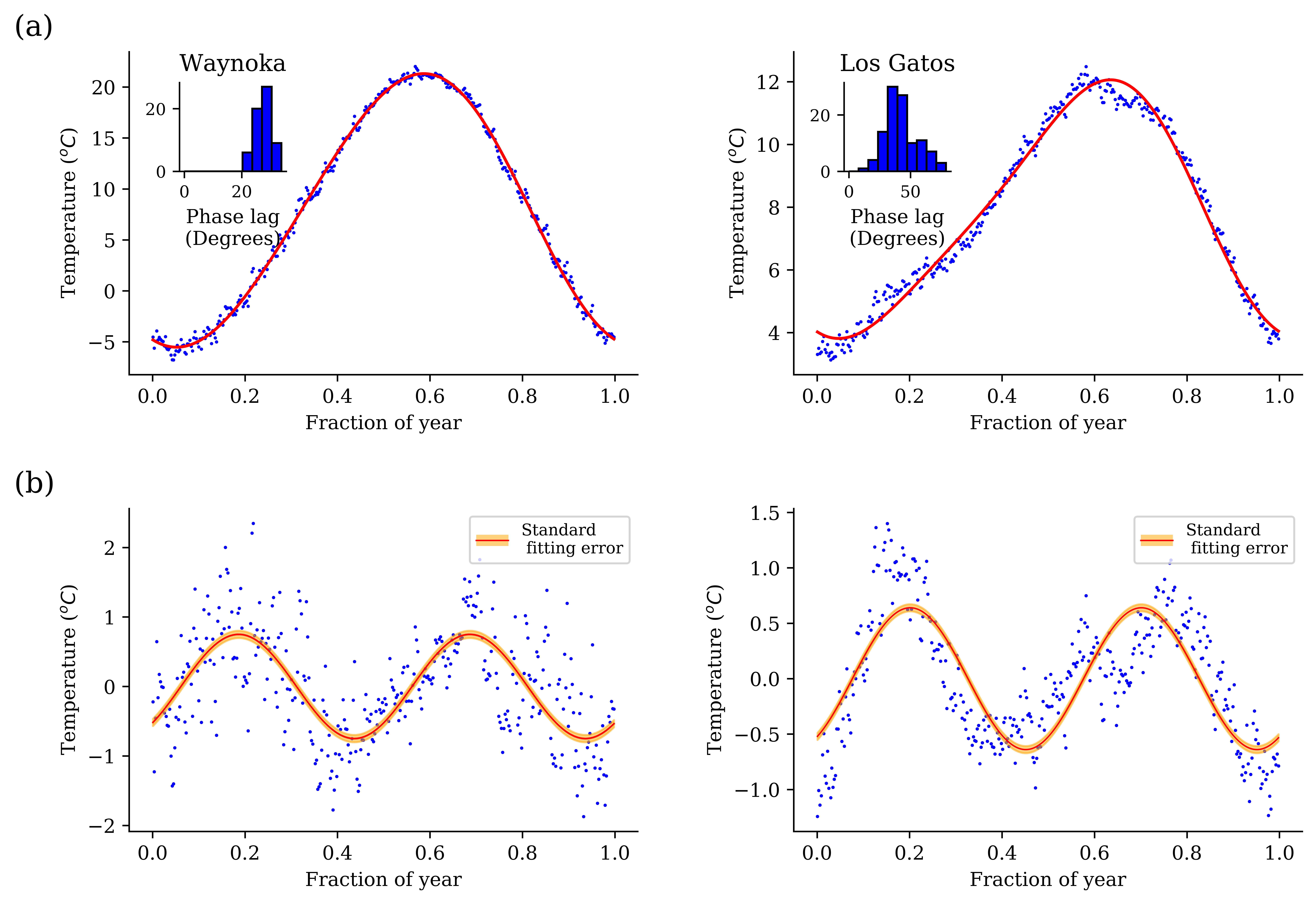}
    \caption{(a) Historically averaged temperature vs. day of year for two local weather stations showing quality of fit to sinusoidal dependence. Left plot is the station with maximal annual gain, marked “L” in Fig. 3; right plot is the station with minimal gain. Horizontal axis is the fraction of year measured from the winter solstice, marked “R” in Fig. 3. Red curve is the five-parameter fit (Eq. 1) including the average temperature and annual and semiannual frequencies. Histograms show individual year phase lags to show consistency over time. (b) Semiannual variation vs. day of year for the same stations. Red curve is the seasonal cycle with the average and the annual component used in (a) subtracted. Orange band gives the error range arising from statistical uncertainties in the fitted parameters, as given in Table 1.}
    \label{Fig. 1}
\end{figure}

\begin{table}
    \centering
    \includegraphics[width=\textwidth]{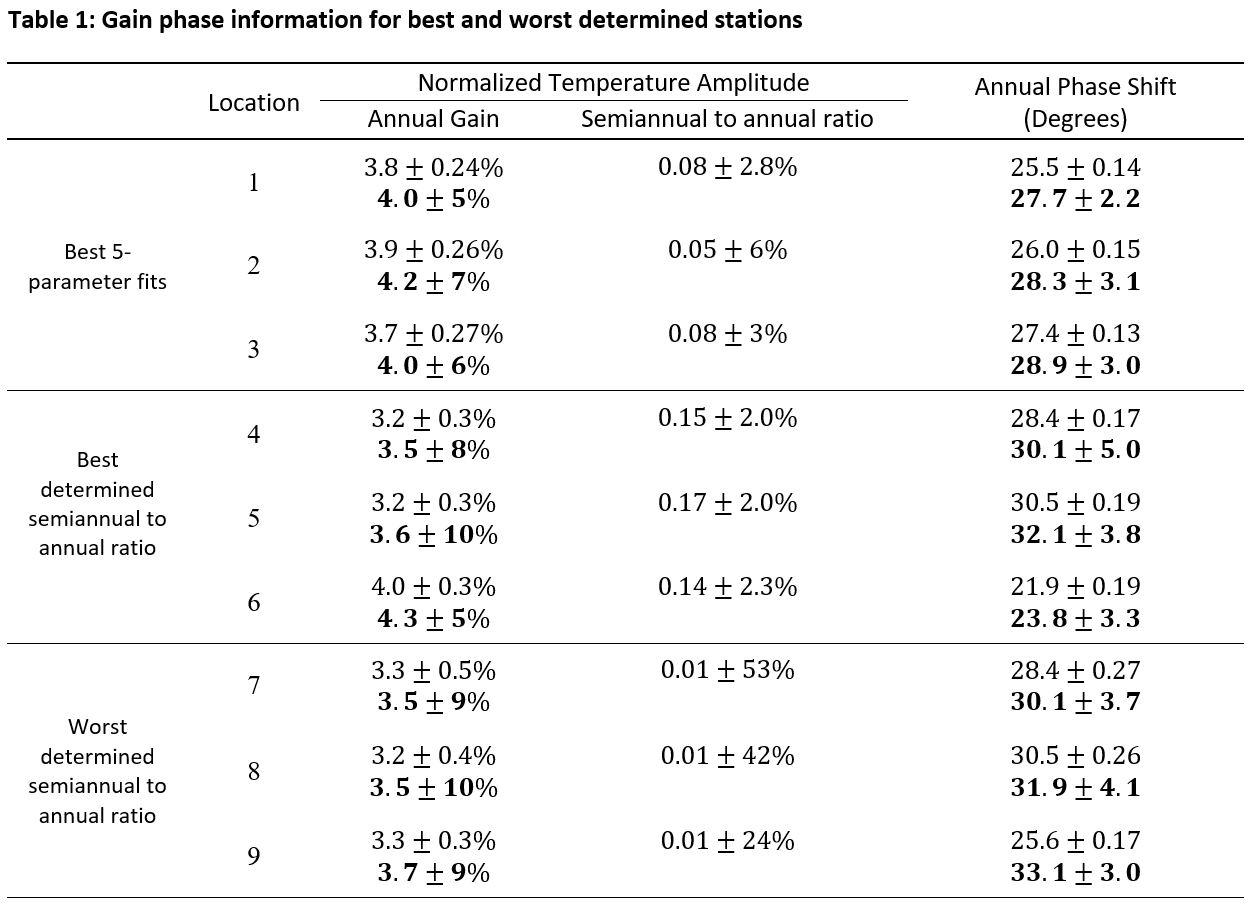}
    \caption{Amplitude and phase information for best and worst determined stations, as identified in the first column. Locations in the second column are indicated on Fig. 3. Annual temperature gains are normalized by the mid-ocean annual gain magnitude defined in the text. The semiannual to annual amplitude ratio is given in the fourth column. Historically averaged quantities with standard fit parameter errors as uncertainty ranges, are given in the upper row for each station. Individual year annual gains and phase shifts, with standard deviations as uncertainty ranges, are given in the bottom row and set in bold. The spread for individual year semiannual quantities are omitted as discussed in Methods. Full information on all stations analyzed is reported in the Supporting Information.}
    \label{Table 1}
\end{table}

\begin{figure}
    \centering
    \includegraphics[width=0.70\columnwidth]{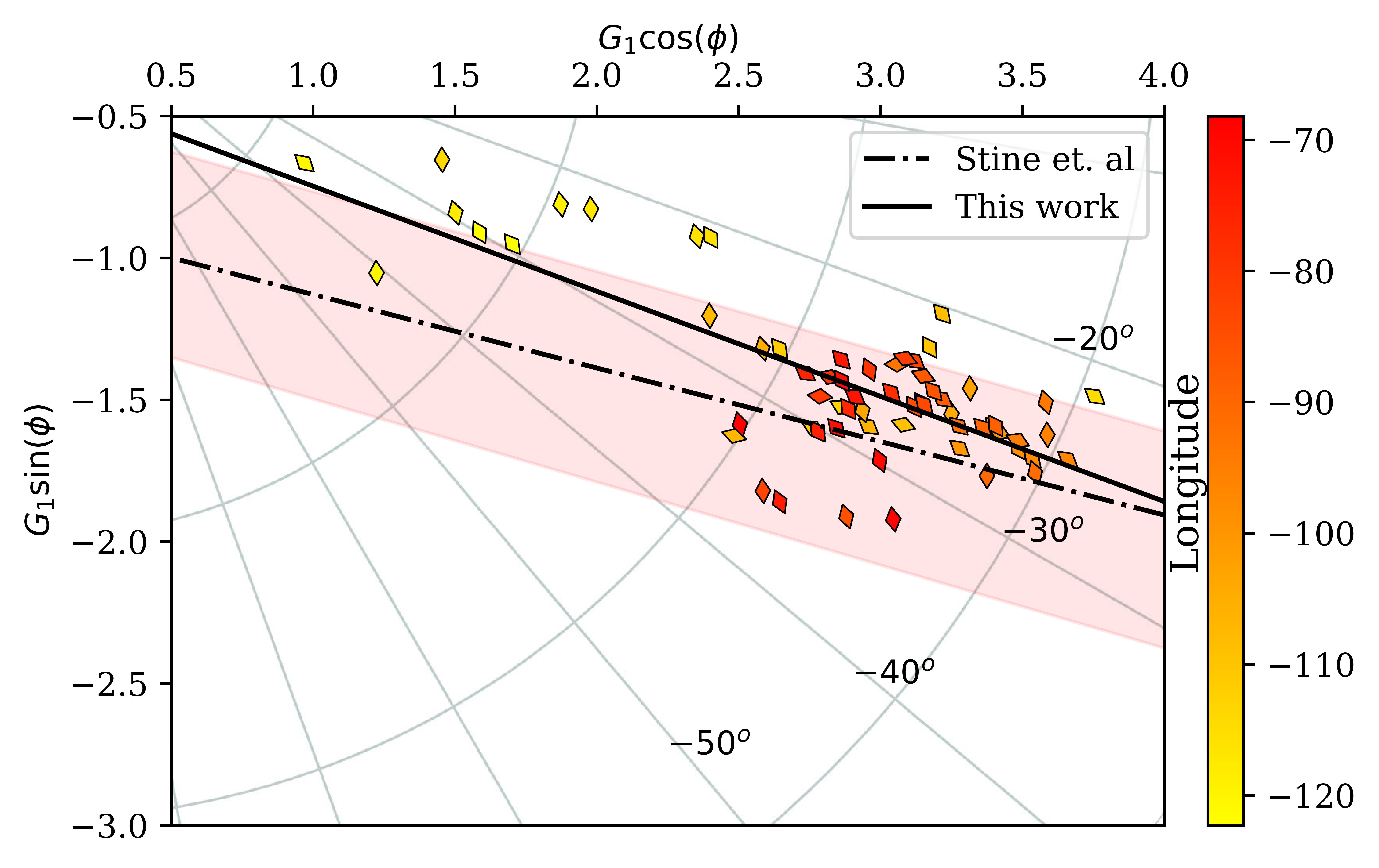}
    \caption{Scatter plot of in-phase ($\cos\phi$) and in-quadrature ($\sin\phi$) gains for all stations in this study. Radial lines and arcs are curves of constant phase and amplitude, respectively. Gains are expressed as multiples of the mid-ocean gain magnitude defined in the text. The solid line is a linear fit to our data. The dashed line and shaded zone indicate the range of data points in the corresponding plot from Stein et al. (2009) Fig. S4b. We obtained the dashed line by drawing a straight line on Fig. S4b that visually represents the trend of the Stine data. The dashed line above is the transcription of this trend line. We obtained the boundaries of the shaded zone by drawing parallel lines on either side of the trend line of Fig S4b so as to encompass about 2/3 of the data points. The shaded zone above is the transcription of this encompassed region.}
    \label{Fig. 2}
\end{figure}

\begin{figure}
    \centering
    \includegraphics[width=0.70\columnwidth]{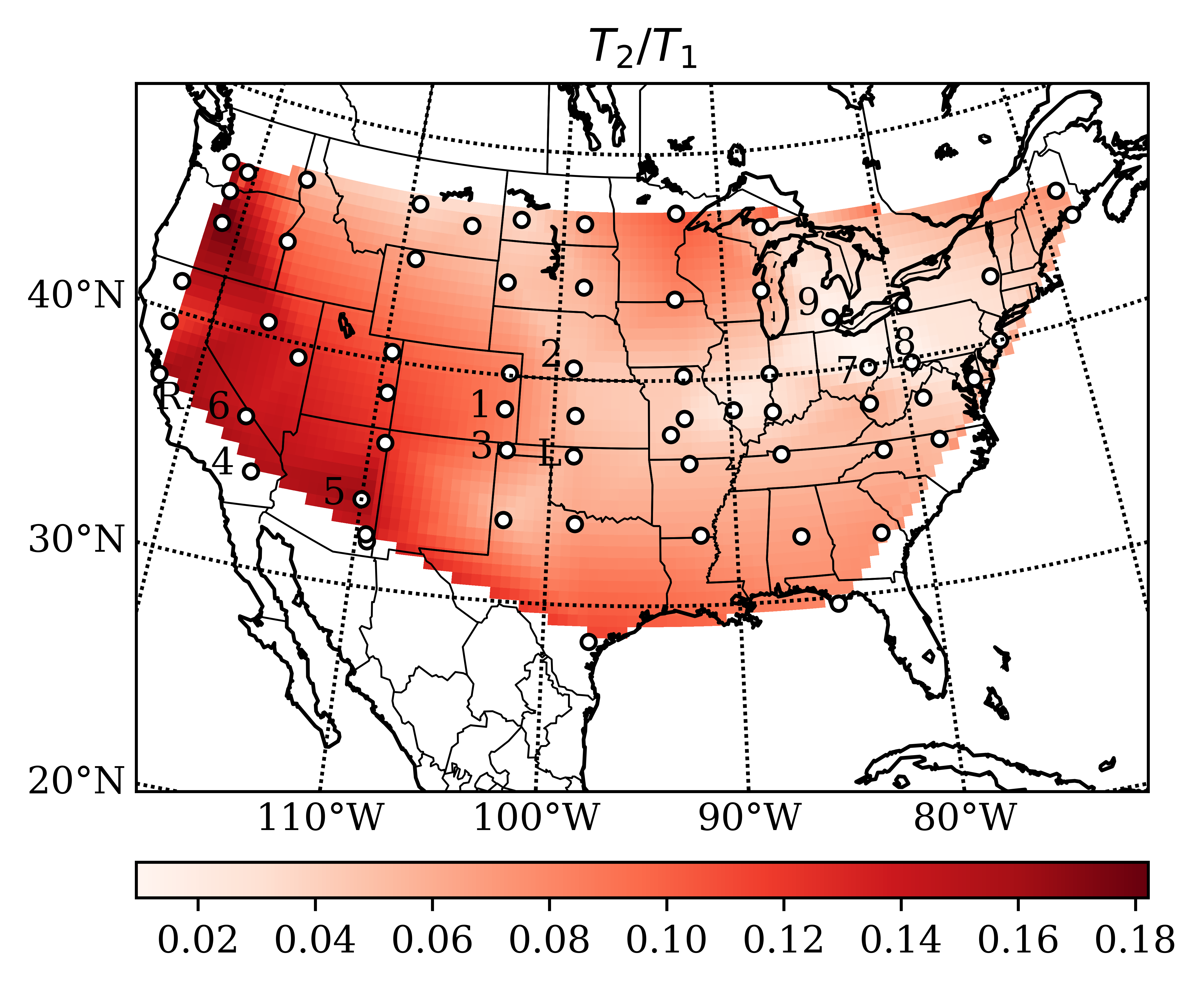}
    \caption{Heat map of the ratio of semiannual to annual temperature amplitude for all stations of this study. White circles are locations of the stations in our selected sample. Stations enumerated in Table 1 are numbered 1 to 9. Station on the left in Fig. 1 is labeled “L”, and the right, labeled “R”. The map was constructed using standard linear interpolation over a mesh-grid of 0.5 by 0.5 resolution \cite{2020SciPy-NMeth}}
    \label{Fig. 3}
\end{figure}

\begin{figure}
    \centering
    \includegraphics[width=0.70\columnwidth]{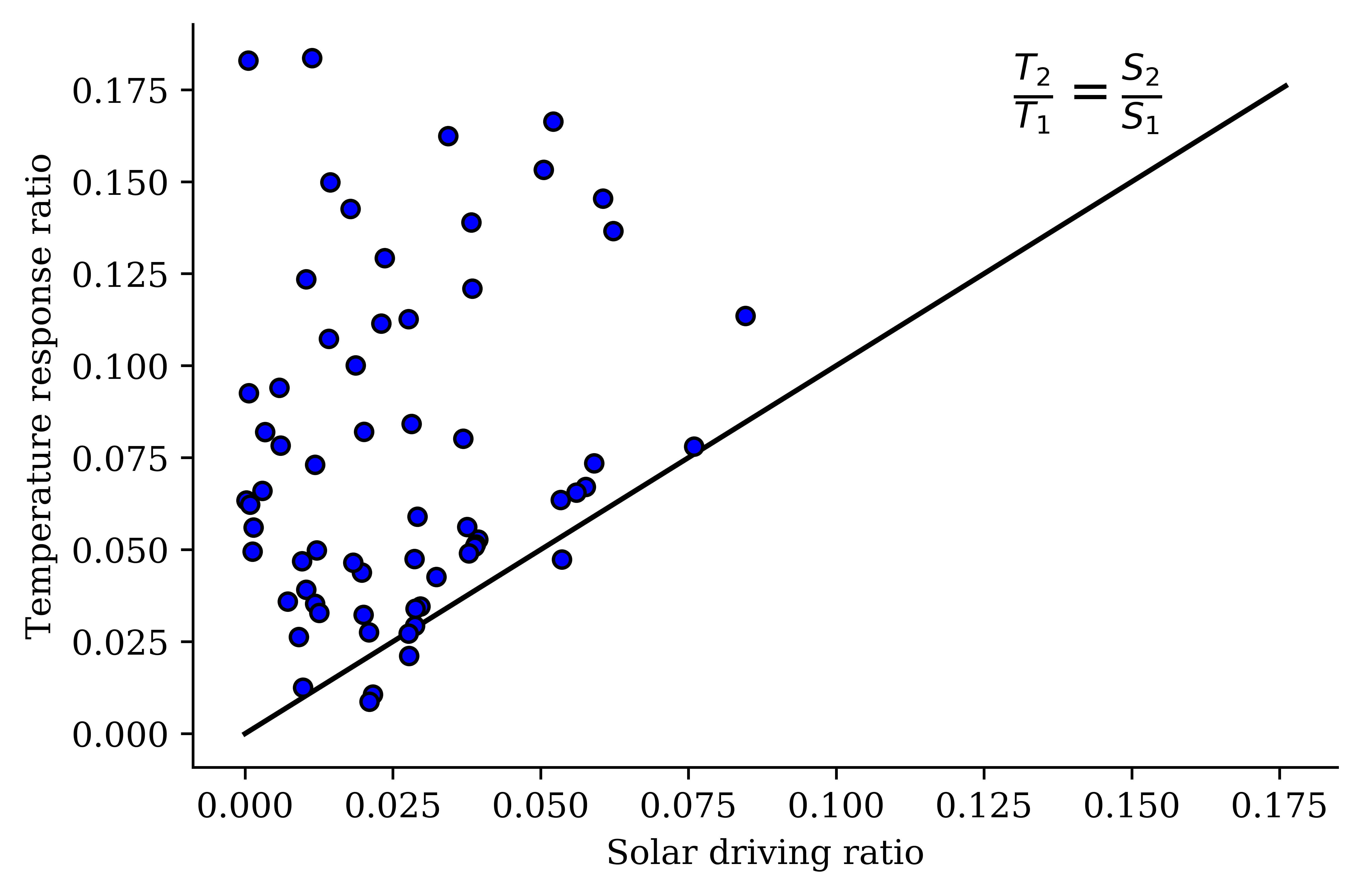}
    \caption{Scatter plot of the ratio of semiannual to annual insolation amplitudes $S_2/S_1$ vs. relative semiannual temperature amplitude $T_2/T_1$ for all weather stations. The lack of weak semiannual responses below the solid line suggests a floor set by the direct semiannual insolation.}
    \label{Fig. 4}
\end{figure}

\begin{figure}
    \centering
    \includegraphics[width=0.70\columnwidth]{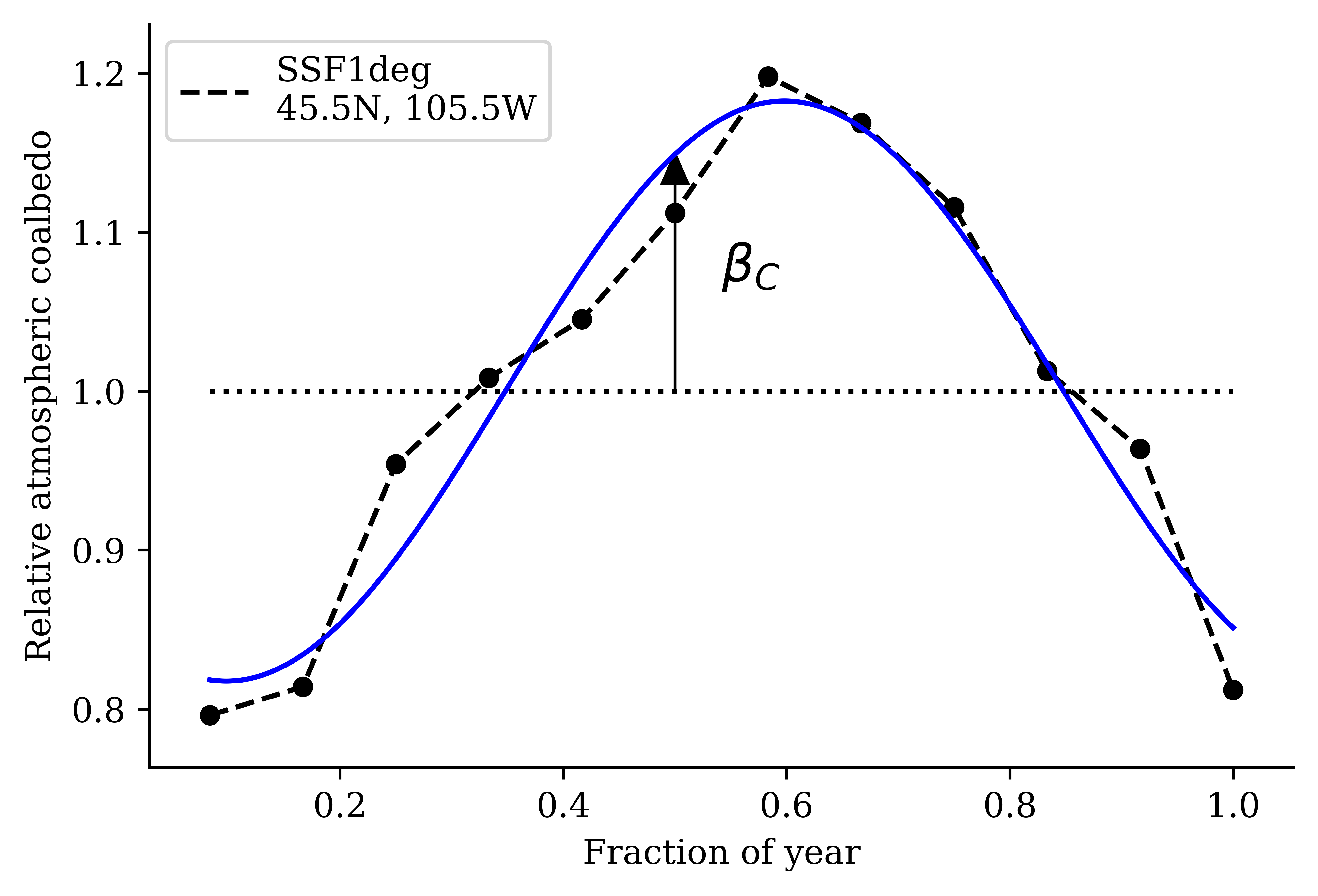}
    \renewcommand{\thefigure}{A1}
    \caption{The annual top-of-atmosphere coalbedo profile for location 45.5N, 105.5W. The Blue curve gives the fit to a sinusoidal function of period a year, with the annual average coalbedo factored out. The arrow spans the amplitude at half a period, marking the dimensionless $\beta_C\approx 0.15$. This falls within the range of our proposed $\beta_C.$}
    \label{Fig. A1}
\end{figure}

\begin{figure}
    \centering
    \includegraphics[width=0.70\columnwidth]{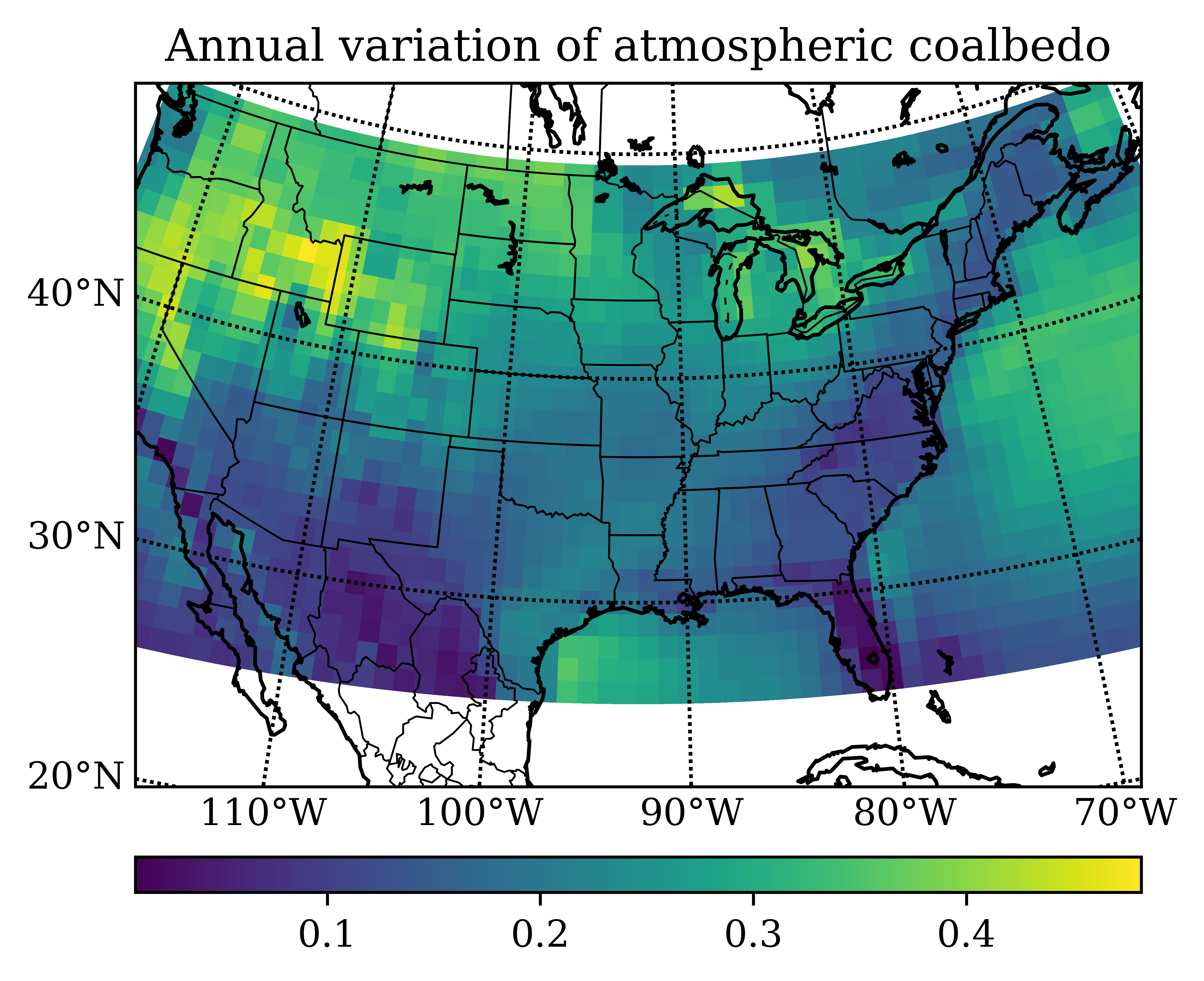}
    \renewcommand{\thefigure}{A2}
    \caption{Heat map of the annual sinusoidal amplitude of top-of-atmosphere coalbedo. Monthly coalbedo data at each location were plotted and fitted to an annual sinusoidal function as in Fig. A1. The amplitudes of the annual component of coalbedo are calculated and plotted at their corresponding location in this map.}
    \label{Fig. A2}
\end{figure}

\end{document}